\documentclass[11pt]{article}

\RequirePackage{fix-cm} 
\usepackage{color}

\usepackage{ntheorem}
\usepackage[T1]{fontenc}

\usepackage{enumerate}
\usepackage{graphicx}
\usepackage{newlfont}
\usepackage{dsfont}
\usepackage{amssymb,amsmath,amsfonts}
\usepackage{epstopdf}
\usepackage{authblk}

\usepackage{geometry}
\newtheorem{definition}{Definition}{}
\newtheorem{corollary}{Corollary}{}
\newtheorem{proposition}{Proposition}{}
\newtheorem{theorem}{Theorem}{}
\newtheorem{remark}{Remark}{}
\newtheorem{lemma}{Lemma}{}

\newcommand{\Real}{{\mathds R}} 

\makeatletter
\newcommand\semiHuge{\@setfontsize\semiHuge{14.5}{14.5}}
\makeatother

\makeatletter
\newcommand{\vast}{\bBigg@{4}}
\newcommand{\Vast}{\bBigg@{5}}
\makeatother

\makeatletter 
\renewcommand{\maketitle}{\bgroup\setlength{\parindent}{0pt}
\begin{flushleft}
  \textbf{\semiHuge  \@title}
  
  \textbf{\large  \@author}
\end{flushleft}\egroup
}
\makeatother

\title{Synchronization in Networks of Diffusively Coupled Nonlinear Systems:\\[1mm] Robustness Against Time-Delays\\[2mm]}
\author{%
Carlos Murguia$^{1}$, Henk Nijmeijer$^{2}$, and Justin Ruths$^{3}$\\
$^{1}$Department of Electrical Engineering, University of Melbourne, Australia\\
    $^{2}$Department of Mechanical Engineering, Eindhoven University of Technology, The Netherlands\\
           $^{3}$Departments of Mechanical and Systems Engineering, University of Texas at Dallas, USA\\\vspace{2mm}
           
    \underline{$^{1}$carlos.murguia@unimelb.edu.au}\\
    \underline{$^{2}$h.nijmeijer@tue.nl }\\
    \underline{$^{3}$jruths@utdallas.edu }
}

\begin{document}
\maketitle

\textbf{Abstract: }In this manuscript, we study the problem of robust synchronization in networks of diffusively time-delayed coupled nonlinear systems. In particular, we prove that, under some mild conditions on the input-output dynamics of the systems and the network topology, there always exists a unimodal region in the parameter space (coupling strength $\gamma$ versus time-delay $\tau$), such that if $\gamma$ and $\tau$ belong to this region, the systems synchronize. Moreover, we show how this unimodal region scales with the network topology, which, in turn, provides useful insights on how to design the network topology to maximize robustness against time-delays. The results are illustrated by extensive simulation experiments of time-delayed coupled Hindmarsh-Rose neural chaotic oscillators.\\


\section{Introduction}

The emergence of synchronization in networks of coupled dynamical systems is a fascinating topic in various scientific disciplines ranging from biology, physics, and chemistry to social networks and technological applications. For instance, in biology, it is well known that thousands of fireflies light up simultaneously \cite{Strog}, and that groups of Japanese tree frogs (Hyla japonica) may show synchronous behavior in their calls \cite{Aihara}. In medicine and neuroscience, clusters of synchronized pacemaker neurons regulate our heartbeat \cite{Peskin}, synchronized neurons in the olfactory bulb allow us to detect and distinguish between odors \cite{Gray}, and our circadian rhythm is synchronized to the 24-h day-night cycle \cite{Czeisler}. In engineering, elements of synchronization are present in power networks \cite{Bullo}, in the velocity of platoons of vehicles \cite{Ploeg}, and in robotics, where multiple robots carry out tasks that cannot be achieved by a single one \cite{Rodriguez}. Several more examples of synchronous behavior in science and engineering can be found in, for instance, (Refs.\hspace{1mm}\cite{Blek,Pikov,Strog,Murguia11,Murguia12,Murguia13}) and references therein.\vspace{1mm}

One of the first technical results regarding synchronization of coupled nonlinear systems is presented in (Refs. \cite{Fujisaka,Pecora}). In these papers, the authors show that coupled chaotic oscillators may synchronize in spite of their high sensitivity to initial conditions. After these results, considerable interest in the notion of synchronization of general nonlinear systems has arisen. Here, we focus on synchronization in networks of identical nonlinear systems interacting through diffusive time-delayed couplings on networks with general topologies. Diffusive time-delayed couplings arise naturally for interconnected systems since the transmission of signals is expected to take some time. Time-delays caused by signal transmission and/or faults in the communication channels affect the behavior of the interconnected systems (e.g., in terms of stability and/or performance).\vspace{1mm}

This manuscript follows the same research line as (Refs. \cite{Pogr1,Erik2,Erik4,Carlos1,Carlos2}), where sufficient conditions for synchronization of diffusively time-delayed coupled \emph{semipassive systems} with and without time-delays are presented. In particular, the authors in Ref. \cite{Erik2} prove that under some mild conditions, there always exists a region $\cal{S}$ in the parameter space (coupling strength $\gamma$ versus time-delay $\tau$), such that if $(\gamma,\tau) \in \cal{S}$ then the systems synchronize. In order to derive their results, the authors assume that the individual systems are \emph{semipassive} \cite{Pogr3} with respect to the coupling variable (the measurable output) and the corresponding internal dynamics has some desired stability properties (\emph{convergent} internal dynamics \cite{Pav}). In the same spirit, here we prove that the region $\cal{S}$ is always bounded by a \emph{unimodal function} $\varphi(\gamma)$ defined on some set $\cal{J} \subset \Real$; and consequently, that there always exists an optimal coupling strength $\gamma^*$ that leads to the maximum time-delay $\tau^*=\varphi(\gamma^*)$ that can be induced to the network without compromising the synchronous behavior. Therefore, for $\gamma = \gamma^*$ and for any $\tau \leq \tau^*$, the systems synchronize, i.e., the gain $\gamma = \gamma^*$ leads to the best tolerance against time-delays of the closed-loop dynamics. Moreover, we analyze the effect of the network topology on the values of both the optimal $\gamma^*$ and the maximum time-delay $\tau^*$, i.e., we show how the eigenvalues of the corresponding Laplacian matrix affect $\varphi(\gamma)$, $\gamma^*$, and $\tau^*$. This, in turn, gives  insights into designing the network topology in order to enhance robustness against time-delays. The results of this manuscript are based on our preliminary conference paper Ref. \cite{MURGUIA201674}.\vspace{1mm}

The remainder of the paper is organized as follows. In Section \ref{sec2}, we recall some important definitions needed for the subsequent sections. The notion of \emph{semipa\-ssivity}, \emph{convergent systems}, and some basic terminology of \emph{graph theory} are introduced. The class of systems under study, the definition of diffusive time-delayed couplings, and the problem formulation are given in Section \ref{sec3}. The main results are presented in Section \ref{sec4}. Simulation experiments of coupled Hindmarsh-Rose neurons are given in Section \ref{sec5} to illustrate our results. Finally, concluding remarks are presented in Section \ref{sec6}.

\section{Preliminaries}\label{sec2}

Throughout this paper, the following notation is used: the symbol $\Real_{>0}$($\Real_{\geq 0}$) denotes the set of positive (nonnegative) real numbers. The Euclidian norm in $\Real^n$ is denoted simply as $|\cdot|$, $|x|^2=x^Tx$, where $^T$ defines transposition. The notation $\text{col}(x_1,...,x_n)$ stands for the column vector composed of the elements $x_1,...,x_n$. This notation will be also used in case the components $x_i$ are vectors. The induced norm of a matrix $A \in \Real^{n \times n}$, denoted by $\| A \|$, is defined as $\| A \| = \max_{x \in \Real^n,|x|=1}|Ax|$. The $n \times n$ identity matrix is denoted by $I_n$ or simply $I$ if no confusion can arise. Likewise, the $n \times m$ matrices composed of only ones and only zeros are denoted as $\mathbf{1}_{n \times m}$ and $\mathbf{0}_{n \times m}$, respectively. The spectrum of a matrix $A$ is denoted by $spec(A)$. For any two matrices $A$ and $B$, the notation $A \otimes B$ (the Kronecker product \cite{Bollobas}) stands for the matrix composed of submatrices $A_{ij}B$ , where $A_{ij}$, $i,j=1,...,n$, stands for the $ij$th entry of the $n \times n$ matrix $A$. Let $\mathcal{X} \subset \Real^n$ and $\mathcal{Y} \subset \Real^m$. The space of continuous functions from $\mathcal{X}$ to $\mathcal{Y}$ is denoted by $\mathcal{C}(\mathcal{X},\mathcal{Y})$. If the functions are (at least) $r \geq 0$ times continuously differentiable, then it is denoted by $\mathcal{C}^r(\mathcal{X},\mathcal{Y})$. If the derivatives of a function of all orders ($r=\infty$) exist\-, the function is called smooth and if the derivatives up to a sufficiently high order exist the function is named sufficiently smooth. For simplicity of notation, we often omit the explicit dependence of time $t$.

\subsection{Communication Graphs}

Given a set of interconnected systems, the communication topology is encoded through a communication graph. The convention is that system $i$ receives information from system $j$ (and viceversa) if and only if there is a link between node $j$ and node $i$ in the communication graph. Let $\mathcal{G} = (\mathcal{V},\mathcal{E},A)$ denote a weighted undirected graph, where $\mathcal{V} = \{v_1,v_2,...,v_k\}$ is the set of nodes, $\mathcal{E} \subseteq \mathcal{V} \times \mathcal{V}$ is the set of edges, and $A$ is the weighted adjacency matrix with nonnegative elements $a_{ij}=a_{ji} \geq 0$. The neighbors of $v_i$ is the set of edges to a node $v_i$ and it is denoted as $\mathcal{E}_i$. If the graph does not contain self-loops, it is called simple. Throughout this manuscript, it is assumed that the communication graph is \emph{strongly connected}, i.e., for every two nodes $(i,j) \in \mathcal{V}$, there is at least one path connecting $i$ and $j$. If two nodes have an edge in common, they are called \emph{adjacent}. Assume that the network consists of $k$ nodes, then the \emph{adjacency matrix} $A \in \Real^{k \times k}:=  a_{ij}$ with $a_{ij}>0$, if $\{i,j\}\in \mathcal{E}$ and $a_{ij}=0$ otherwise. Finally, we introduce the \emph{degree matrix} $D \in \Real^{k \times k}:= \text{diag}\{d_1,...,d_k\}$ with $d_i=\sum_{j \in \mathcal{E}_{i}}a_{ij}$, and $L := D - A$, which is called the \emph{Laplacian matrix} of the graph $\cal{G}$, see Ref. \cite{Bollobas} for further details.

\subsection{Semipassive Systems}
Consider the system
\begin{subequations}\label{1DF1}
\begin{eqnarray}
\dot{x} &=& f(x,u),  \label{1P} \\
y &=& h(x), \label{1PP}
\end{eqnarray}
\end{subequations}
with state $x \in \Real^n$, input $u \in \Real^m$, output $y \in \Real^m$, sufficiently smooth functi\-ons $f:\Real^n \times \Real^m  \rightarrow \Real^n$, and $h:\Real^n \rightarrow \Real^m$.

\begin{definition}{\emph{[Ref. \cite{Pogr3}]}.}
The dynamical system \eqref{1DF1} is called $\mathcal{C}^r$-semipassive if there exists a nonnegative storage function $V \in \mathcal{C}^r(\Real^n,\Real_{ \geq 0})$ such that the differential inequality $\dot{V}(x,u)\leq y^{T}u-H(x)$ is satisfied, where the function $H \in \mathcal{C}(\Real^n,\Real)$ is nonnegative outside some ball, i.e., $\exists \text{ } \varphi >0\text{ } \text{s.t.} \text{ } \left\vert x\right\vert \geq \varphi \rightarrow H(x)\geq \varrho (\left\vert x\right\vert)$, for some continuous nonnegative  function $\varrho(\cdot)$ defined for $\left\vert x\right\vert \geq \varphi$. If the function $H(\cdot)$ is positive definite outside some ball, then the system \eqref{1DF1} is said to be strictly $\mathcal{C}^r$-semipassive.
\end{definition}

\begin{remark}
System \eqref{1DF1} is $\mathcal{C}^r$-passive (strictly $\mathcal{C}^r$-passive) if it is $\mathcal{C}^r$-semipassive (strictly $\mathcal{C}^r$-semipassive) with $H(\cdot)$ being positive semidefinite (positive definite).
\end{remark}

In light of Remark 1, a (strictly) $\mathcal{C}^r$-semipassive system behaves like a (strictly) passive system \cite{Willems_1} for sufficiently large $|x|$. The concept of semipassivity allows us to find simple conditions which ensure bounded trajectories of the interconnected systems. The class of strictly semipassive systems includes, e.g., the chaotic Lorenz system \cite{Pogr1}, and many models that describe the action potential dynamics of individual neurons \cite{Erik3}.

\subsection{Convergent Systems}
Consider the system (\ref{1P}) and assume that $f(\cdot)$ is Lipschitz in $x$, $u(\cdot)$ is piecewise continuous in $t$ and takes values in some compact set $u \in U \subseteq \Real^m$.
\begin{definition}
System \eqref{1P} is said to be \emph{convergent} if and only if for any bounded signal $u(t)$ defined on the whole interval $(-\infty,+\infty)$ there is a unique bounded globally asympto\-tically stable solution $\bar{x}_u(t)$ defined in the same interval for which it holds that, $\lim_{t\rightarrow \infty }\left\vert x(t)-\bar{x}_u(t)\right\vert =0$ for all initial conditions.
\end{definition}
For a \emph{convergent system}, the limit solution is solely de\-termined by the external excitation $u(t)$ and not by the initial condition. A sufficient condition for a system to be convergent is obtained in Ref. \cite{Demi} and later extended in Ref. \cite{Pav} is presented in the following proposition.

\begin{proposition}{\emph{[Refs. \cite{Demi} and \cite{Pav}]}.}
If there exists a po\-sitive definite symmetric matrix $P \in \Real^{n \times n}$ such that all the eigenvalues $\lambda_i(Q)$ of the symmetric matrix
\begin{equation}
Q(x ,u)=\frac{1}{2} \left( P\left( \frac{\partial f}{\partial x }%
(x ,u)\right) +\left( \frac{\partial f}{\partial x }(x
,u)\right) ^{T} P \right),   \label{7P}
\end{equation}
are negative and separated from zero, i.e., there exists a constant $c \in \Real_{>0}$ such that $\lambda_i (Q)\leq -c <0,$ for all  $i \in \{1,...,n \}$, $u \in U$, and $x \in \Real^n$, then system \eqref{1P} is globally exponentially convergent. Moreover, for any pair of solutions $x_1(t),x_2(t) \in \Real^n$ of \eqref{1P}, the following is satisfied
\[
\frac{d}{dt}\Big(\big(x_{1}-x_{2}\big)^TP\big(x_{1}-x_{2}\big)\Big)\leq -\alpha \left\vert
x_{1}-x_{2}\right\vert ^{2},
\]
with constant $\alpha := \frac{c}{\lambda_{\max}(P)}$ and $\lambda_{\max}(P)$ being the largest eigenvalue of the symmetric matrix $P$.
\end{proposition}

\section{System Description and Problem Statement}\label{sec3}

Consider $k$ identical nonlinear systems of the form\\
\begin{align}
\dot{\zeta}_{i} &= q(\zeta _{i},y_{i}),  \label{1} \\
\dot{y}_{i} &= a(\zeta _{i},y_{i})+CB u_i,  \label{2}
\end{align}\\
with $i \in \mathcal{I}:=\{1,...,k\}$, state $x_i := \text{col}(\zeta_i,y_i) \in \Real^n$, internal state $\zeta_i \in \Real^{n-m}$, output $y_i \in \Real^m$, input $u_i \in \Real^m$, sufficiently smooth functions $q:\Real^{n-m} \times \Real^m  \rightarrow \Real^{n-m}$ and $a:\Real^{n-m} \times \Real^m  \rightarrow \Real^m$, matrices $C \in \Real^{m \times n}$ and $B \in \Real^{n \times m}$, and the matrix $CB \in \Real^{m \times m}$ being similar to a positive definite matrix. For the sake of simplicity, it is assumed that $CB = I_m$ (results for the general case with $CB$ being similar to a positive definite matrix can be easily derived). It is assumed that (\ref{1})-(\ref{2}) is \emph{strictly $\cal{C}^1$-semipassive} and the \emph{internal dynamics} \cite{Henk2} (\ref{1}) is an \emph{exponentially convergent system}. Let the $k$ systems (\ref{1})-(\ref{2}) interact on simple strongly connected graph through the \emph{diffusive time-delayed coupling}\vspace{2mm}
\begin{equation}
u_{i}(t) = \gamma \sum_{j\in \mathcal{E}_{i}}a_{ij}\left( y_{j}(t-\tau
)-y_{i}(t-\tau )\right) ,  \label{3}
\end{equation}\\
where $\tau \in \Real_{ \geq 0}$ denotes a constant time-delay, $y_j(t-\tau)$  and $y_i(t-\tau)$ are the time-delayed outputs of the  $j$-th and $i$-th systems, $\gamma \in \Real_{\geq 0}$ denotes the coupling strength, $a_{ij} \geq 0$ are the weights of the interconnections, and $\mathcal{E}_{i}$ is the set of neighbors of system $i$. It is assumed that the graph is \emph{undirected}, i.e., $a_{ij}=a_{ji}$. Moreover, since the coupling strength is encompassed in the constant $\gamma$, it is assumed without loss of generality that $\text{max}_{i \in \cal{I} }\sum_{j \in \mathcal{E}_{i}} a_{ij} = 1$. Note that all signals in coupling (\ref{3}) are time-delayed. Such a coupling may arise, for instance, when the systems are interconnected through a centralized control law. Coupling (\ref{3}) can be written in matrix form as follows\vspace{1mm}
\begin{equation}
u = -\gamma \left( L\otimes I_{m}\right) y(t - \tau),  \label{3Matrix}\vspace{1mm}
\end{equation}
with $u:=\text{col}(u_{1},\ldots,u_{k})$, $y:=\text{col}(y_{1},\ldots,y_{k})$, and Laplacian matrix $L = L^T \in \Real^{k \times k}$. The authors in Ref. \cite{Erik2} prove that the $k$ coupled systems (\ref{1})-(\ref{3}) asymptotically synchronize provided that $\gamma$ is sufficiently large and the product of the coupling strength and the time-delay $\gamma \tau$ is sufficiently small. It follows that there exists a region $\cal{S}$ in the parameter space, such that if $(\gamma,\tau) \in \cal{S}$, the systems synchronize. In this manuscript, we go one step further by showing that this region $\cal{S}$ is actually bounded by a \emph{unimodal function} $\varphi: \mathcal{J} \to \Real_{\geq 0}$, $\gamma \mapsto \varphi(\gamma)$. Therefore, there  exists an optimal coupling strength $\gamma^*$ that leads to the maximum delay $\tau^*=\varphi(\gamma^*)$ that can be induced without compromising the synchronous behavior. Moreover, we characterize the function $\varphi(\gamma)$, the optimal $\gamma^*$, and the maximum time-delay $\tau^*$ in terms of the spectrum of the Laplacian matrix $L$ and the vector fields $q(\cdot)$ and $a(\cdot)$.

\section{Main Results}\label{sec4}

In this section, we give sufficient conditions for synchronization in networks of the diffusively time-delayed coupled semipassive systems. However, before we start thinking about synchronization, it is necessary to ensure that the solutions of the closed-loop system (\ref{1})-(\ref{3}) are well defined, i.e., the solutions exist and are bounded. In the following lemma (adapted from Ref. \cite{Erik2}), we give sufficient conditions for \emph{ultimate boundedness} \cite{Kha2002} of the solutions of the coupled systems interacting on \emph{simple strongly connected graphs}.

\begin{figure}[t]
\begin{center}
\includegraphics[width=6.25cm]{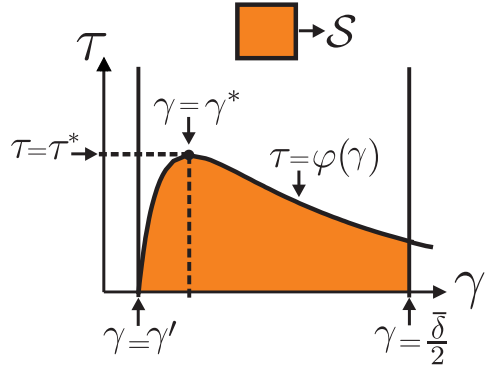}
\caption{Synchronization Region $\mathcal{S}$.}
\end{center}
\label{Fig1}
\end{figure}

\newpage

\begin{lemma}
Consider $k$ coupled systems \eqref{1}-\eqref{3} with coupling strength $\gamma \in \Real_{ \geq 0}$ and time-delay $\tau \in \Real_{ \geq 0}$ on a simple undirected strongly connected graph $\cal{G}$. Assume that:\\[1.5mm]
\emph{\textbf{(H4.1)}} Each system \eqref{1}-\eqref{2} is strictly $\mathcal{C}^1$-semipassive with input $u_i$, output $y_i$, radially unbounded storage function $V(x_i)$, and the functions $H(x_i)$ are such that there exists a constant $R \in \Real_{>0}$ such that $|x_i|>R$ implies that $H(x_i) - \delta |y_i|^2 > 0$ for some constant $\delta \in \Real_{>0}$. \\[1.5mm]
Let $\bar{\delta}$ be the largest $\delta$ that satisfies \emph{(H4.1)}. Then, the solutions of the coupled systems \eqref{1}-\eqref{3} are ultimately bounded for any finite $\tau \in \Real_{ > 0}$ and $\gamma < \bar{\delta}/2$.
\end{lemma}
\emph{\textbf{Proof}}: By assumption, each system (\ref{1})-(\ref{2}) is strictly $\mathcal{C}^1$-semipassive with input $u_i$, output $y_i$, and radially unbounded function $V(x_i)$. Define the functional
\begin{equation}
W(x_{t}(\theta )):=\sum_{i=1}^{k}\nu _{i}\Big( V(x_{i})+ \gamma \sum\limits_{j\in \mathcal{E}%
_{i}}a_{ij}\int\limits_{-\tau }^{0}\left\vert y_{i}(t+s)\right\vert
^{2}ds\Big),\notag
\end{equation}
where $x = \text{col}(x_1,...,x_k)$, $x_t(\theta) = x(t + \theta) \in \cal{C}$, $ \theta \in [-\tau,0]$, $\mathcal{C} = [-\tau,0] \rightarrow \Real^{kn}$ being the Banach space of continuous functions mapping the interval $[-\tau,0]$ into $\Real^{kn}$, and the constants $\nu_i$ denoting the entries of the left eigenvector corresponding to the simple zero eigenvalue of the Laplacian matrix $L$, i.e., $\nu  =\left( \nu _{1},...,\nu _{k}\right) ^{T}$ and $\nu ^{T}L =\nu ^{T}\left(D-A\right) =0$. Note that $L$ is singular by construction and since it is assumed that the graph is strongly connected, then the zero eigenvalue is simple. Using the Perron-Frobenius theorem, it can be shown that the vector $\nu$ has strictly positive real entries, i.e., $\nu_i>0$ for all $i \in \mathcal{I}$, see Ref. \cite{Bollobas}.  Then, by assumption
\begin{equation}
\dot{W} \leq \sum_{i=1}^{k}\nu _{i}\Big( y_{i}^Tu_{i}-H(x_{i})+\gamma
\sum\limits_{j\in \mathcal{E}_{i}}a_{ij}\left( \left\vert y_{i}\right\vert
^{2}-\left\vert y_{i}^{\tau }\right\vert ^{2}\right) \Big).  \notag
\end{equation}
Consider the term
\begin{equation}
\sum\limits_{i=1}^{k}\nu _{i}y_{i}^{T}u_{i}=\gamma
\sum\limits_{i=1}^{k}\sum\limits_{j\in \mathcal{E}_{i}}\nu
_{i}a_{ij}y_{i}^{T}(y^{\tau}_{j}- y^{\tau}_{i}),  \label{PT1_1}
\end{equation}
using Young's inequality, it follows that
\begin{equation*}
\sum\limits_{i=1}^{k}\nu _{i}y_{i}^{T}u_{i}\leq \frac{\gamma }{2}%
\sum\limits_{i=1}^{k}\sum\limits_{j\in \mathcal{E}_{i}}\nu _{i}a_{ij}\Big(%
2|y_{i}|^{2}+|y_{i}^{\tau }|^{2}+|y_{j}^{\tau }|^{2}\Big).
\end{equation*}
Combining the previous results and using the fact that $\nu ^{T}L =0$, and $\max_{i\in \cal{I}}\sum_{j\in \mathcal{E}_{i}}a_{ij}=1$, the following is satisfied $\dot{W}\leq \sum_{i=1}^{k}\nu _{i}( -H(x_{i})+2\gamma %
\left\vert y_{i}\right\vert ^{2})$. By assumption (H4.1), there exist positive constants $R,\delta \in \Real_{>0}$ such that $|x_i|>R$ implies that $H(x_i) - \delta |y_i|^2 > 0$. Let $\bar{\delta}$ be the largest $\delta$ that satisfies (H4.1) for arbitrarily large $R<\infty$. Then, for $\gamma$ satisfying $2\gamma \leq \bar{\delta}$ and for sufficiently large $|x|$, it follows that $\dot{W}<0$. By construction, the functional $W$ is lower and upper bounded by some $\cal{K}_{\infty}$-functions \cite{Kha2002}. Hence, there exists a constant $\sigma \in \Real_{>0}$ such that $\dot{W}<0$ for $\sigma$ and $x$ satisfying ${W}(x) \geq \sigma$. Then, solutions starting in the set $\{W(x) \leq \sigma \}$ will remain there for future time since $\dot{W}$ is negative on the boundary $W(x)= \sigma$. Moreover, for any $x$ in the set $\{ W(x) \geq \sigma^*\}$ with $\sigma^* > \sigma$, the function $\dot{W}(x)$ is strictly negative, which implies that, in this set, $W(x)$ decreases monotonically until the solutions enter the set $\{ W(x) \leq \sigma\}$ again; and therefore, we can conclude that the solutions of the closed-loop system (\ref{1})-(\ref{3}) exist and are ultimately bounded for any finite $\tau \geq 0$ and $\gamma \leq \frac{\bar{\delta}}{2}$. \hfill $\blacksquare$

\begin{remark}
The result stated in Lemma 1 is independent of
the time-delay. Therefore, if the conditions stated in Lemma 1 are satisfied, the solutions of the closed-loop system \emph{(\ref{1})-(\ref{3})} are ultimately bounded for arbitrary large time-delays.
\end{remark}

Next, we give sufficient conditions for global synchronization of the coupled systems. Define the stacked state  $x:= \text{col}(x_1,\ldots,x_k)$ ($x_i := \text{col}(\zeta_i,y_i)$) and the \emph{synchronization manifold} $\mathcal{M} := \{x \in \Real^{kn}\text{ } |\hspace{.5mm} x_i = x_j, \hspace{.5mm} \forall \text{ \ } i,j \in \cal{I} \}$. The coupled systems (\ref{1})-(\ref{3}) are said to synchronize if the synchronization manifold $\mathcal{M}$ is invariant under the closed-loop dynamics and contains an asympto\-tically stable subset. Clearly, if the systems are interconnected through (\ref{3}), the coupling functions vanish on $\mathcal{M}$; hence, the manifold $\cal{M}$ is positively invariant under the dynamics (\ref{1})-(\ref{3}), see Ref. \cite{Erik4} for details. In the following theorem, we give  conditions for the existence of an asympto\-tically stable subset of the synchronization manifold. In particular, we prove that under some mild assumptions, there exists a region $\cal{S}$ bounded by a \emph{unimodal function} such that if $(\gamma,\tau) \in \cal{S}$, the systems synchronize.

\begin{definition}
A function $\varphi: \mathcal{J} \to \Real_{\geq 0}$, $\gamma \mapsto \varphi(\gamma)$ is called \emph{unimodal} if for some value $\gamma^* \in \cal{J}$, it is monotonically increasing for $\gamma \leq \gamma^*$ and monotonically decreasing for $\gamma \geq \gamma^*$. Hence, the maximum value of $\varphi(\gamma)$ is given by $\varphi(\gamma^*)$ and there are no other maxima.\vspace{1mm}
\end{definition}

\begin{theorem} Consider $k$ coupled systems \emph{(\ref{1})-(\ref{3})} with coupling strength $\gamma \in \Real_{ \geq 0}$ and time-delay $\tau \in \Real_{ \geq 0}$ on a simple undirected strongly connected graph $\cal{G}$. Let the conditions of Lemma 1 be satisfied and assume that:\\[1mm]
\emph{\textbf{(H4.2)}} The internal dynamics \eqref{1} is an exponentially convergent system, i.e., there is a positive definite matrix $P$ such that the eigenvalues of the symmetric matrix
\begin{equation}
\frac{1}{2} \left( P\left( \frac{\partial q}{\partial \zeta_i }%
(\zeta_i,y_i)\right) + \left( \frac{\partial q}{\partial \zeta_i }(\zeta_i
,y_i)\right) ^{T}P \right),   \label{7P}
\end{equation}
are uniformly negative and bounded away from zero for all $\zeta_i \in \Real^{n-m}$ and $y_i \in \Real^m$.\\[1mm]
Then, there exist a constant $\gamma^\prime \in \Real_{>0}$ and a unimodal functi\-on $\varphi:\mathcal{J}:=[\gamma^\prime,\infty) \rightarrow \Real_{\geq 0}$, $\gamma \mapsto \varphi(\gamma)$, where $\varphi(\gamma^\prime)= 0$ and $\lim_{\gamma \to \infty} \varphi(\gamma)=0$, such that if $(\gamma,\tau) \in \mathcal{S}:= \{\gamma,\tau \in \Real_{ \geq 0} \text{ }|\text{ } \gamma < \frac{\bar{\delta}}{2},\text{ } \gamma > \gamma^{\prime},\text{ } \tau < \varphi(\gamma) \}$, then there exists a globally asympto\-tically stable subset of the synchronization manifold $\mathcal{M}$.
\end{theorem}
The proof of Theorem 1 can be found in the appendix. The result stated in Theorem 1 amounts to the following. The solutions of (\ref{1})-(\ref{3}) are ultimately bounded and the systems asymptotically synchronize provided that the coupling strength $\gamma$ is sufficiently large and the time-delay $\tau$ is smaller than some \emph{unimodal function} $\varphi(\gamma)$, see Figure 1. Therefore, there exists a region $\cal{S}$ (colored area in Figure 1) such that if $(\gamma,\tau) \in \cal{S}$, there exists a globally asympto\-tically stable subset of the synchronization manifold $\cal{M}$. Estimates of both the constant $\gamma^\prime$ and the \emph{unimodal function} $\varphi(\gamma)$ are derived in the proof of Theorem 1, in (\ref{condpred}) and (\ref{neww2}), respectively. They depend on the network topology (more specifically on the eigenvalues of the Laplacian matrix), the dynamics of the individual subsystems, i.e., the vector fields $q(\cdot)$ and $a(\cdot)$, and the bounds on the solutions of the closed-loop system. Note that our results are meant to prove \emph{existence} of the synchronization region $\cal{S}$ and to provide a \emph{qualitative analysis} of the unimodal bound $\varphi(\gamma)$; the estimates of $\gamma^\prime$ and $\varphi(\gamma)$ given in appendix may be conservative.

\begin{figure*}[t]
\includegraphics[width=16.75cm]{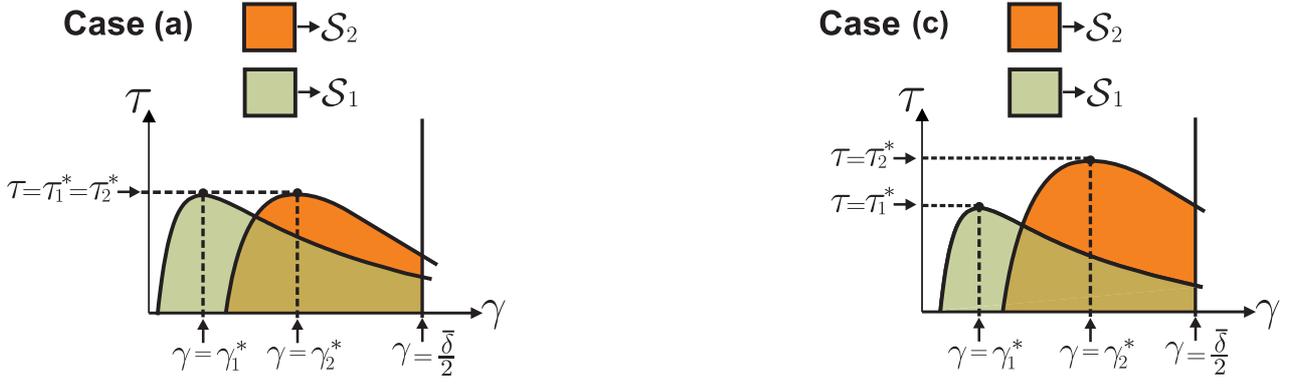}
\caption{Synchronization Regions $\mathcal{S}_s$, $s=1,2$. Left: Case (a) in Corollary 2. Right: Case (c) in Corollary 2.}
\label{Fig2}
\end{figure*}

\begin{corollary}
Consider $k$ coupled systems \eqref{1}-\eqref{3} with coupling strength $\gamma \in \Real_{ \geq 0}$ and time-delay $\tau \in \Real_{ \geq 0}$ on a simple undirected strongly connected graph $\cal{G}$. Assume that the conditions stated in Theorem 1 are satisfied. Then, for every $\bar{\gamma} \in \cal{J}$, there exists a maximum time-delay $\bar{\tau} \in \Real_{ \geq 0}$, such that if $\gamma = \bar{\gamma}$, the systems synchronize for all $\tau \leq \bar{\tau}$. Moreover, there exists an optimal coupling strength $\gamma = \gamma^* \in \cal{J}$ and its corresponding maximum time-delay $\tau^{\ast}:= \max(\tau \in \mathcal{S})$ such that $\bar{\tau} < \tau^{\ast}$ for all $\gamma \neq \gamma^*$.\\[1mm]
\emph{\textbf{\emph{Proof:}} The assertion follows from unimodality of the function $\varphi(\gamma)$.}
\end{corollary}

The result stated in Corollary 1 implies that for every strongly connected graph $\cal{G}$, there exists an optimal coupling strength $\gamma^*$ that leads to the maximum time-delay $\tau^*=\varphi(\gamma^*)=\max(\tau \in \mathcal{S})$ that can be induced to the network without compromising the synchronous behavior, i.e., there exists a gain $\gamma = \gamma^*$ that leads to the best tolerance against time-delays of the closed-loop system, see Figure 1. In the following corollary, we characterize the effect of the network topology on the values of both the optimal $\gamma^*$ and the maximum time-delay $\tau^*$, i.e., we study the effect of the eigenvalues of the Laplacian matrix $L$ on the unimodal function $\varphi(\gamma)$.\vspace{1mm}

Let $\bar{\mathcal{G}}:= \{ \mathcal{G}_s \hspace{1mm} | \hspace{1mm} s \in \mathbb{N} \hspace{1mm}\}$ denote the set of all simple strongly connected undirected graphs. For each graph $\mathcal{G}_s$, consider $k=k_s$ coupled systems (\ref{1})-(\ref{3}) with coupling strength $\gamma=\gamma_s \in \Real_{ \geq 0}$ and delay $\tau=\tau_s \in \Real_{ \geq 0}$ interacting on $\mathcal{G}_s$. Assume that the conditions stated in Theorem 1 are satisfied, then from Theorem 1, there exist constants $\gamma_s^\prime \in \Real_{>0}$ and unimodal functi\-ons $\varphi_s:\mathcal{J}_s \rightarrow \Real_{\geq 0}$, $s \in \mathbb{N}$, such that if $(\gamma_s,\tau_s) \in \mathcal{S}_s$, where
\[
\mathcal{S}_s:=\{\gamma_s,\tau_s \in \Real_{ \geq 0} \text{ }|\text{ } \gamma_s < \bar{\delta}/2,\text{ } \gamma_s > \gamma_s^{\prime},\text{ } \tau_s < \varphi_s(\gamma_s) \},
\]
the $k_s$ systems of each $\mathcal{G}_s$ asympto\-tically synchronize. Let $\gamma_s^*$ and $\tau_s^*=\max(\tau_s \in \mathcal{S}_s)$ be the optimal coupling strength and the corresponding maximum time-delay of $\mathcal{G}_s$ and define $\bar{\tau}^*:= \{ \tau_s^* \hspace{1mm} | \hspace{1mm} s \in \mathbb{N} \hspace{1mm}\}$, i.e., the set of all maximum time-delays of $\bar{\mathcal{G}}$, and
\begin{equation}
\tau _{\max }^{\ast }:=\max (\bar{\tau}^{\ast }).  \label{tmax}
\end{equation}
\begin{corollary}
For $s=1,2$. Consider $k=k_s$ interconnected systems \eqref{1}-\eqref{3} with coupling strength $\gamma=\gamma_s \in \Real_{ \geq 0}$ and time-delay $\tau=\tau_s \in \Real_{ \geq 0}$ on a simple undirected strongly connected graph $\mathcal{G}_s$, i.e., two independent diffusively time-delayed coupled networks. Assume that $\mathcal{G}_1 \neq \mathcal{G}_2$ and the conditions stated in Theorem 1 are satisfied. Then, by Theorem 1, there exists a region $\mathcal{S}_s$ such that if $(\gamma_s,\tau_s) \in \mathcal{S}_s$ the $k_s$ systems of $\mathcal{G}_s$ asympto\-tically synchronize. Let $\gamma_s^*$ and $\tau_s^*=\max(\tau_s \in \mathcal{S}_s)$ be the optimal coupling strength and the corresponding maximum time-delay of $\mathcal{G}_s$. Additionally, let $L_s \in \Real^{k_s \times k_s}$ be the Laplacian matrix of $\mathcal{G}_s$ with real positive eigenvalues $\lambda_j(L_s)$, $j = 2,\ldots,k_s$, and $\lambda_2(L_s) \leq \ldots \leq \lambda_{k_s}(L_{s})$. Then:\\[1mm]
\emph{\textbf{(a)}} If  $\frac{\lambda_{k_1}}{\lambda_2}(L_1) = \frac{\text{ } \lambda_{k_2}}{\lambda_2}(L_2) > 1$, then $\tau_1^* = \tau_2^*$. In addition, $\lambda_2(L_1) \geq \lambda_2(L_2)$ implies $\gamma_1^* \leq \gamma_2^*$.\\[1mm]
\emph{\textbf{(b)}} If $\frac{ \lambda_{k_1}}{\lambda_2}(L_1) = \frac{\text{ } \lambda_{k_2}}{\lambda_2}(L_2) = 1$, then $\tau_1^* = \tau_2^* = \tau_{\max}^*$. In addition, $\lambda_2(L_1) \geq \lambda_2(L_2)$ implies $\gamma_1^* \leq \gamma_2^*$.\\[1mm]
\emph{\textbf{(c)}} If $\frac{ \lambda_{k_1}}{\lambda_2}(L_1) > \frac{\text{ } \lambda_{k_2}}{\lambda_2}(L_2)$, then $\tau_1^* < \tau_2^*$. In addition, $\lambda_2(L_1) \geq \lambda_2(L_2)$ implies $\gamma_1^* \leq \gamma_2^*$.
\end{corollary}
The proof of Corollary 2 can be found in the appendix. The result stated in Corollary 2 amounts to the following. The effect of the network topology on the maximum time-delay $\tau^*$ is solely determined by the quotient $(\frac{\lambda_{k}}{\lambda_2})$. Case (a): implies that any two strongly connected networks with the same quotient $(\lambda_{k}/\lambda_2)$ have the same maximum time-delay $\tau^*$. Additionally, in this case, the value of the optimal coupling strength $\gamma^*$ is also determined by $\lambda_2$, i.e., if the two networks have the same quotient, then the larger the $\lambda_2$ the smaller the $\gamma^*$, and vice versa. Case (b): implies that networks with quotient equal to one have the best tolerance against time-delays, i.e., if $\frac{\lambda_{k}}{\lambda_2}=1$, then $\tau^* = \tau^*_{\max}$ with $\tau^*_{\max}$ as defined in (\ref{tmax}). Finally, case (c) implies that the larger the quotient $\frac{\lambda_k}{\lambda_2}$ the smaller the $\tau^*$, and vice versa. See Figure 2.

\section{Simulation Experiment}\label{sec5}
\textbf{A. Network Topology, Strict Semipassivity, and  Convergence.} Consider a network of $k_s$, $s \in \{1,\ldots,7\}$, systems coupled according to the graphs $\mathcal{G}_s$ depicted in Figure 3. For each network, the weights of the interconnections are set to $a_{ij} = 1/k_s$ if $\{i,j\} \in \mathcal{E}_s$ and $a_{ij} = 0$ otherwise. The networks are strongly connected and undirected. Each system in the networks is assumed to be a Hindmarsh-Rose neuron\cite{HindMarsh}, of the form
\begin{equation}
\left\{
\begin{array}{l}
\dot{\zeta}_{1i}=1-5y_{i}^{2}-\zeta_{1i}, \\[1mm]
\dot{\zeta}_{2i}=0.005(4y_{i}+6.472-\zeta_{2i}), \label{EX1} \\[1mm]
\dot{y}_{i} \hspace{1.5mm}=-y_{i}^{3}+3y_{i}^{2}+\zeta_{1i}-\zeta_{2i}+3.25+u_{i},%
\end{array}%
\right.
\end{equation}
with output $y_i \in \Real$, internal states $\zeta_{i1},\zeta_{i2} \in \Real$, state $x_i = \text{col}(\zeta_{i1},\zeta_{i2},y_{i}) \in \Real^3$, input $u_i \in \Real$,  and $i \in \mathcal{I} = \{1,\ldots,k_s\}$. It is well known that the Hindmarsh-Rose neuron (\ref{EX1}) has a chaotic attractor for $u_i=0$, see Ref. \cite{HindMarsh}. Furthermore, in Ref. \cite{Erik3}, the authors prove that the Hindmarsh-Rose neuron is strictly $\mathcal{C}^1$-semipassive with storage function $V(\zeta_{1i},\zeta_{2i},y_{i}):=\frac{1}{2}y_{i}^{2}+\sigma \zeta_{1i}^{2}+25\zeta_{2i}^{2}$, constants $\varsigma_1,\varsigma_2 \in (0,1)$, $0<\sigma<\frac{4\varsigma_1(1-\varsigma_2)}{25}$, and
\begin{align}
&H(\zeta_{1i},\zeta_{2i},y_{i}) = \varsigma _{1}y_{i}^{4}-3y_{i}^{3}-\frac{1}{%
4\sigma (1-\varsigma _{2})}y_{i}^{2}  \notag \\
&+\left( \sigma \varsigma _{2}-\frac{25\sigma ^{2}}{4(1-\varsigma _{1})}%
\right) \zeta_{1i}^{2}+\frac{1}{4}\zeta_{2i}^{2}-1.618\zeta_{2i}  \notag \\
&+\sigma (1-\varsigma _{2})\left( \zeta_{1i}-\frac{1}{2\sigma (1-\varsigma _{2})%
}y_{i}\right) ^{2}-\sigma \zeta_{1i}  \notag \\
&+(1-\varsigma _{1})\left( y_{i}^{2}+\frac{5\sigma }{2(1-\varsigma _{1})}%
\zeta_{1i}\right) ^{2} -3.25y_{i}.  \label{Ex2}
\end{align}
Moreover, the ($\zeta_{1i},\zeta_{2i}$)-dynamics (the internal dynamics) is \emph{exponentially convergent} (in the sense of Definition 2), i.e., it satisfies the Demidovich condition (\ref{7P}) with $P = I_2$; hence, assumption (H4.2) in Theorem 1 is satisfied. This particular experiment is taken from Ref. \cite{Neefs}, where a detailed experimental study is presented.\\[2mm]
\textbf{B. Bounded Solutions and Synchronization.}
It can be easily verified that the function $H(x_{i})$ in (\ref{Ex2}) satisfies the boundedness assumption (H4.1) for arbitrary large coupling strength $\gamma$. Therefore, by Lemma 1, the solutions of the coupled systems (\ref{EX1}),(\ref{3}) with $\gamma=\gamma_s$ and $\tau=\tau_s$ are ultimately bounded for any finite coupling strength $\gamma_s$ and time-delay $\tau_s$. Finally, given that all the graphs in Figure 3 are strongly connected and (H4.1) and (H4.2) are satisfied, then, by Theorem 1, there exist regions $\mathcal{S}_s$, $s=1,\ldots,7$, (as depicted in Figure 1), such that if $(\gamma_s,\tau_s) \in \mathcal{S}_s$, the systems synchronize.\\[3mm]
\textbf{C. Simulation Results.} In Figure 4, we show results obtained through extensive computer simulations. We depict the synchronization regions $\mathcal{S}_s$, $s=1,\ldots,7$ for each network topology $\mathcal{G}_s$. These regions are clearly bounded by unimodal functions; and therefore, for each network, there exists an optimal coupling strength  $\gamma^*_s$ and its corres\-ponding maximum time-delay $\tau^*_s$ that can be induced to the network without compromising the synchronous behavior. These maximum time-delays are strongly influenced by the network topology, (see the proofs of Theorem 1 and Corollary 2). In Table 1, we show the numerical values of the optimal coupling strengths, the maximum time-delays, and the quotients $\lambda_{k_s}/\lambda_2$. These values match with the theoretical predictions given in Corollary 2.

\begin{figure}[t]
\centering
\includegraphics[width=5.0cm]{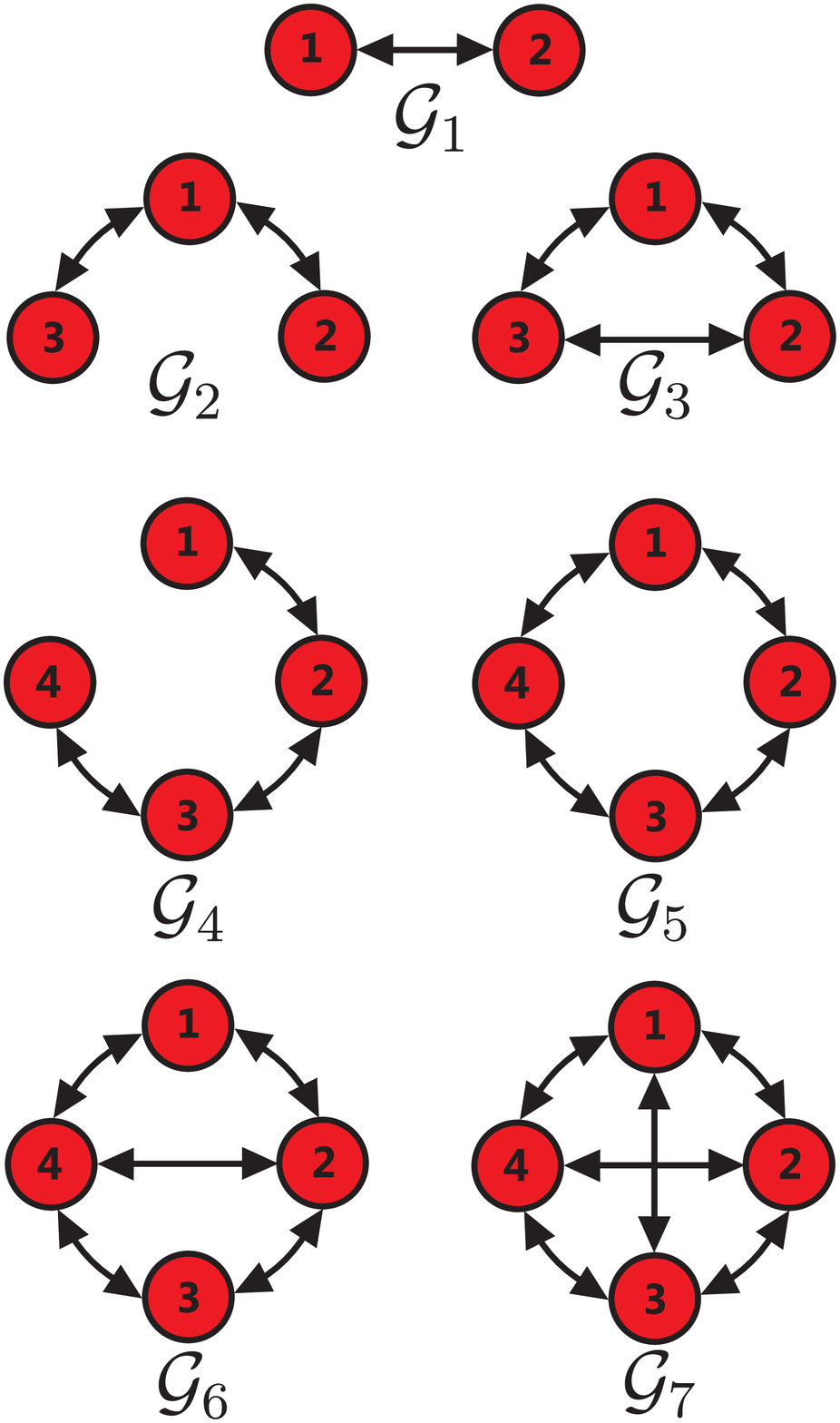}
\caption{Network topologies.}
\label{Fig3}
\end{figure}

\begin{figure}[t]
\centering
\includegraphics[width=7.0cm]{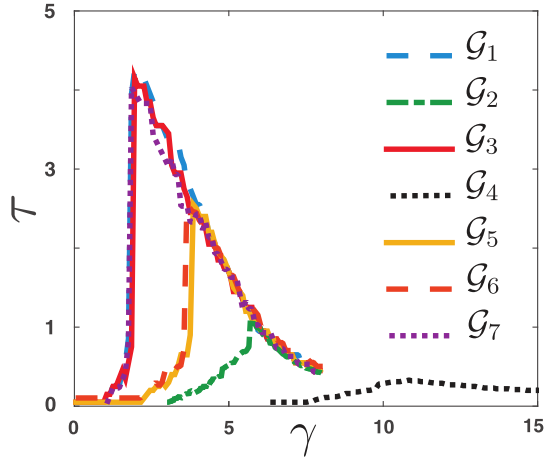}
\caption{Synchronization regions $\mathcal{S}_s$ computed for diffe\-rent topologies $\mathcal{G}_s$, $s=1,\ldots,7$. See Ref. \cite{Neefs} for comparison with experimental results.}
\label{Fig3}
\end{figure} 

\section{Conclusions}\label{sec6}

We have presented a result on network synchronization of coupled nonlinear systems in the case when the coupling functions are subject to constant time-delays. Using the notions of semipassivity and convergent systems, we have provided sufficient conditions which guarantee ultimate boundedness of the solutions of the coupled systems and (global) state synchronization. In particular, we have proved that, under some mild assumptions, there always exists a region $\mathcal{S}$ in the parameter space (coupling strength $\gamma$ versus time-delay $\tau$), such that if $\gamma,\tau \in \mathcal{S}$, the systems synchronize. We proved that this region $\cal{S}$ is always bounded by a \emph{unimodal function} $\varphi(\gamma)$; and consequently, that there always exists an optimal coupling strength $\gamma^*$ which leads to the maximum time-delay $\tau^*=\varphi(\gamma^*)$ that the network can tolerate without breaking the synchrony. In Corollary 2, we have provided tools for selecting the network topology in order to enhance robustness against time-delays of the coupled systems. Case (b) in Corollary 2 implies that networks with quotient $(\lambda_k/\lambda_2)$ equal to 1 have the best tolerance against time-delays. This is the case for \emph{all-to-all} networks. Finally, we have presented a simulation example using networks of Hindmarsh-Rose neurons to illustrate the results.

\begin{table}[t]
\label{table1}
\caption{Simulation results: Optimal coupling strength $\gamma^*_s$, maximum time-delay $\tau^*_s$, eigen\-values $\lambda_{k_s}$ and $\lambda_2$ of the corresponding Laplacian matrices $L_s$, and their quotient $(\lambda_{k_s}/\lambda_2)$.}
\begin{tabular}{|l|l|l|l|l|l|}
  \hline
   & $\gamma^*_s$ & $\tau^*_s$[ms] & $\lambda_{k_s}$ & $\lambda_2$ & $\lambda_{k_s}/\lambda_2$ \\
  \hline
  $\mathcal{G}_1$ & 2.00& 4.25 & $1$ & $1$ & $1$ \\
  $\mathcal{G}_2$ & 5.70 & 1.10 & $1$ & $\frac{1}{3}$ & $3$ \\
  $\mathcal{G}_3$ & 1.95 & 4.25 & $1$ & $1$ & $1$ \\
  $\mathcal{G}_4$ & 10.6 & 0.33 & $0.8536$ & $0.1464$ & $5.8306$ \\
  $\mathcal{G}_5$ & 3.85 & 2.55 & $1$ & $\frac{1}{2}$ & 2 \\
  $\mathcal{G}_6$ & 3.75 & 2.50 & $1$ & $\frac{1}{2}$ & 2 \\
  $\mathcal{G}_7$ & 1.90 & 4.15 & $1$ & $1$ & $1$ \\
  \hline
\end{tabular}
\centering
\end{table}

\appendix

\section{Proofs} 

\subsection{Proof of Theorem 1}

Let $\zeta = \text{col}(\zeta_1,\ldots,\zeta_k) \in \Real^{k(n-m)}$ and $y = \text{col}(y_1,\ldots,y_k) \in \Real^{km}$. Define $M \in \Real^{(k-1) \times k}$ as
\begin{equation}
M :=\left(
\begin{array}{cc}
\mathbf{1}_{k-1} & -I_{k-1}%
\end{array}%
\right),
\end{equation}
Introduce the set of coor\-dinates $\tilde{\zeta} = (M \otimes I_{n-m})\zeta$ and  $\tilde{y} = (M \otimes I_{m})y$. Note that, $\tilde{y}_1=y_1-y_2,\ldots,\tilde{y}_{k-1}=y_1-y_k$ and $\tilde{\zeta}_1=\zeta_1-\zeta_2,\ldots,\tilde{\zeta}_{k-1}=\zeta_1-\zeta_k$. Then, it follows that $\tilde{y}=\tilde{\zeta}=0$ implies that the systems are synchronized. Assumption (H4.2), Proposition 1, smoothness of the vector fields, and boundedness of the solutions imply the existence of a positive definite function $V_2: \Real^{(k-1)(n-m)} \rightarrow \Real_{\geq 0}$, $\tilde{\zeta} \mapsto V_{2}(\tilde{\zeta})$ such that
\begin{equation}
\dot{V}_{2}(\tilde{\zeta},\tilde{y}) \leq
- \alpha|\tilde{\zeta}| ^{2} + c_{0}\tilde{|\zeta|} \left\vert \tilde{y}\right\vert ,
\label{PPP2}
\end{equation}
for some constants $\alpha,c_0 \in \Real_{>0}$, see Section 5 in Ref. \cite{Pogr1} for further details. Note that
\begin{equation}
\resizebox{.425 \textwidth}{!}
{
$
\tilde{M}= \left(
\begin{array}{ll}
1 & \mathbf{0}_{k-1} \\
\mathbf{1}_{k-1} & -I_{k-1}%
\end{array}%
\right) \Rightarrow  \tilde{M}L\tilde{M}^{-1}=\left(
\begin{array}{cc}
0\text{ \ \ \ } & \mathbf{\ast } \\
\mathbf{0}_{k-1} & \tilde{L}%
\end{array}%
\right) ,
\label{Ltilde}
$
}
\end{equation}
where $L$ denotes the Laplacian matrix. By assumption, the communication graph is \emph{strongly connected} and \emph{undirected}. Then, the Laplacian matrix is symmetric and its eigenvalues are real. Moreover, the matrix $L$ has an algebraically simple eigenvalue $\lambda_1=0$ and $\mathbf{1}_k$ is the correspon\-ding eigenvector \cite{Bollobas}. Since $\text{spec}(\tilde{L})=\text{spec}(L) \backslash \{0\}$, then from Gerschgorin's disc theorem, it can be concluded that the eigenvalues of $\tilde{L}$ are positive real, i.e., the matrix $\tilde{L}$ has eigenvalues $\lambda_2,...,\lambda_k \in \Real_{>0}$ with $0 < \lambda_2 \leq\cdots \leq \lambda_k$. Coupling (\ref{3}) can be written in matrix form as follows
\begin{align}
u(t) &= -\gamma \left( L\otimes I_{m}\right) y(t-\tau),  \label{PP3}
\end{align}
where $u=\text{col}(u_1,...,u_k) \in \Real^{km}$. Denote $\tilde{u}=\text{col}((u_1-u_2),...,(u_1-u_k))$; then
\begin{equation}
\tilde{u}(t)=-\gamma ( \tilde{L}\otimes I_{m}) \tilde{y}(t-\tau),
\label{PP4}
\end{equation}
with $\tilde{L}$ as in (\ref{Ltilde}). In the new coordinates, the closed-loop system can be written as
\begin{align}
\dot{\tilde{\zeta}} &=\tilde{q}(\tilde{y},\tilde{\zeta},y_{1},\zeta _{1}),
\label{PP51} \\
\dot{\tilde{y}} &=\tilde{a}(\tilde{y},\tilde{\zeta},y_{1},\zeta
_{1})-\gamma ( \tilde{L}\otimes I_{m}) \tilde{y}(t-\tau),\label{PP5}
\end{align}
where
\begin{equation}
\resizebox{.425 \textwidth}{!}
{
$
\tilde{a}(\tilde{y},\tilde{\zeta},y_{1},\zeta _{1})=\left(
\begin{array}{c}
a(y_{1},\zeta _{1})-a(y_{1}-\tilde{y}_{1},\zeta _{1}-\tilde{\zeta}_{1}) \\
\vdots  \\
a(y_{1},\zeta _{1})-a(y_{1}-\tilde{y}_{k-1},\zeta _{1}-\tilde{\zeta}_{k-1})%
\end{array}%
\right),  \label{PP6}
$
}
\end{equation}
and
\begin{equation}
\resizebox{.425 \textwidth}{!}
{
$
\tilde{q}(\tilde{y},\tilde{\zeta},y_{1},\zeta _{1})=\left(
\begin{array}{c}
q(y_{1},\zeta _{1})-q(y_{1}-\tilde{y}_{1},\zeta _{1}-\tilde{\zeta}_{1}) \\
\vdots  \\
q(y_{1},\zeta _{1})-q(y_{1}-\tilde{y}_{k-1},\zeta _{1}-\tilde{\zeta}_{k-1})%
\end{array}%
\right) \label{PP6b}.
$
}
\end{equation}
Using Leibniz's rule and continuity of the solutions, the variable $\tilde{y}(t-\tau)$ can be written as
\begin{equation}
\tilde{y}(t-\tau) = \tilde{y}(t) - \int_{-\tau }^{0} \dot{\tilde{y}}(t+s)ds.
\label{91AT3}
\end{equation}
It follows that the dynamics (\ref{PP5}) can be written as
\begin{equation}
\resizebox{.485 \textwidth}{!}
{
$
\dot{\tilde{y}}=a(\tilde{y},\tilde{\zeta},y_{1},\zeta _{1})-\gamma (\tilde{L}%
\otimes I_{m})\tilde{y}+\gamma (\tilde{L}\otimes I_{m})\int_{-\tau }^{0}\dot{%
\tilde{y}}(t+s)ds,  \label{93A}
$
}
\end{equation}
substitution of (\ref{PP5}) in (\ref{93A}) yields
\begin{align}
\dot{\tilde{y}}& =a(\tilde{y},\tilde{\zeta},y_{1},\zeta _{1})-\gamma (\tilde{%
L}\otimes I_{m})\tilde{y}  \label{93AA} \\[1mm]
& -\gamma ^{2}(\tilde{L}^{2}\otimes I_{m})\int_{-\tau }^{0}\tilde{y}(t-\tau
+s)ds  \notag \\
& +\gamma (\tilde{L}\otimes I_{m})\int_{-\tau }^{0}a(\tilde{y},\tilde{\zeta}%
,y_{1},\zeta _{1})(t+s)ds.  \notag
\end{align}
The matrix $\tilde{L}$ is nonsingular and symmetric, then there exists a transformation matrix $U \in \Real^{(k-1)\times(k-1)}$ such that $\| U \|=1$ and $U \tilde{L} U^{-1} = \Lambda$, where $\Lambda$ denotes a diagonal matrix with the nonzero eigen\-values of $L$ on its diagonal. Introduce the change of coordinates $\bar{y}=(U \otimes I_m)\tilde{y}$ and for consistency of notation $\bar{\zeta}=\tilde{\zeta}$. In the new coordinates, the closed-loop system can be written as
\begin{align}
\dot{\bar{\zeta}}& =\bar{q}(\bar{y},\bar{\zeta},y_{1},\zeta _{1}),
\label{ppp1} \\
\dot{\bar{y}}& =\bar{a}(\bar{y},\bar{\zeta},y_{1},\zeta _{1})-\gamma \left(
\Lambda \otimes I_{m}\right) \bar{y}(t)  \label{ppp2} \\[1mm]
& -\gamma ^{2}(\Lambda ^{2}\otimes I_{m})\int_{-\tau }^{0}\bar{y}(t-\tau
+s)ds  \notag \\
& +\gamma (\Lambda \otimes I_{m})\int_{-\tau }^{0}\bar{a}(\bar{y},\bar{\zeta}%
,y_{1},\zeta _{1})(t+s)ds,  \notag
\end{align}
where $\bar{q}(\bar{y},\bar{\zeta},y_{1},\zeta _{1}) := \tilde{q}((U^{-1} \otimes I_m)\bar{y},\bar{\zeta},y_{1},\zeta _{1})$, $\bar{a}(\bar{y},\bar{\zeta},y_{1},\zeta _{1}) := (U \otimes I_m) \tilde{a}((U^{-1} \otimes I_m)\bar{y},\bar{\zeta},y_{1},\zeta _{1})$. Notice that $\bar{y}=\bar{\zeta}=0$ implies that the systems are synchronized because $U$ is nonsingular. Since stability is invariant under a change of coordinates and $\|U\|=1$, then from (\ref{PPP2}), there exists a positive definite function $\bar{V}_2: \Real^{(k-1)(n-m)} \rightarrow \Real_{\geq 0}$, $\bar{\zeta} \mapsto \bar{V}_{2}(\bar{\zeta})$ such that
\begin{equation}
\dot{\bar{V}}_{2}(\bar{\zeta},\bar{y}) \leq
- \alpha|\bar{\zeta}| ^{2} + c_{0}\bar{|\zeta|} \left\vert \bar{y}\right\vert ,
\label{PP2}
\end{equation}
for some constants $\alpha,c_0 \in \Real_{>0}$. Consider the function $V_{3}(\bar{y})=\frac{1}{2}\bar{y}^{T} \bar{y}$. Then
\begin{align}
\dot{V}_{3} \leq &-\gamma \lambda _{2}|\bar{y}|^{2}+\bar{y}^{T}\bar{a}(\bar{%
y},\bar{\zeta},y_{1},\zeta _{1}) \notag\\ &+\gamma \bar{y}^{T}(\Lambda \otimes
I_{m})\int_{-\tau }^{0}\bar{a}(\bar{y},\bar{\zeta},y_{1},\zeta _{1})(t+s)ds
\label{PP7} \notag \\
&-\gamma ^{2}\bar{y}^{T}(\Lambda ^{2}\otimes I_{m})\int_{-\tau }^{0}\bar{y}%
(t-\tau +s)ds.
\end{align}
Ultimate boundedness of the solutions and smoothness of the function $a(\cdot)$ imply that
\begin{align}
\bar{y}^{T} \bar{a}(\bar{y},\bar{\zeta},y_{1},\zeta
_{1})& \leq c_{1}\left\vert \bar{y}\right\vert ^{2}+c_{2}\left\vert \bar{%
y}\right\vert |\bar{\zeta}|,  \notag
\end{align}
for some positive constants $c_1,c_2 \in \Real_{>0}$. Let the function $\mathcal{V}_1(\bar{\zeta},\bar{y}) := V_2(\bar{\zeta}) + V_3(\bar{y})$ be a Lyapunov-Razumikhin function such that if $\mathcal{V}_1(\bar{\zeta}(t),\bar{y}(t))>\mathcal{\varkappa}^{2} \mathcal{V}_1(\bar{\zeta}(t+\theta),\bar{y}(t+\theta))$ for $\theta \in \lbrack -2\tau ,0]$ and some constant $\varkappa>1$, then
\begin{eqnarray}
\mathcal{\dot{V}}_{1} &\leq &-\alpha \left\vert \bar{\zeta}\right\vert
^{2}+\left( \varkappa \tau \lambda _{k}^{2}\gamma ^{2}+c_{1}(1+\varkappa
\gamma \tau \lambda _{k})-\gamma \lambda _{2}\right) \left\vert \bar{y}%
\right\vert ^{2}  \notag \\
&&+\left( c_{0}+c_{2}(1+\varkappa \gamma \tau \lambda _{k})\right)
\left\vert \bar{\zeta}\right\vert \left\vert \bar{y}\right\vert.  \label{102AT3}
\end{eqnarray}
The constant $\varkappa$ can be arbitrarily close to one as long as it is greater than one. Then, for the sake of simplicity, we take $\varkappa$ on the boundary $\varkappa=1$ for the rest of the analysis. Some straightforward algebra shows that (\ref{102AT3}) is negative definite if
\begin{equation}
\Big(\lambda _{2}\gamma -\gamma ^{\prime }\Big)-\lambda _{k}\Big(\lambda
_{k}\gamma +\frac{\bar{c}_{1}}{\bar{c}_{2}}\Big)\gamma \tau -\frac{\lambda
_{k}^{2}}{2\bar{c}_{2}}(\gamma \tau )^{2}>0,  \label{condpred2}
\end{equation}
with
\begin{align}
\gamma ^{\prime }& :=  \frac{\left( c_{0}+c_{2}\right) ^{2}}{4\alpha }+c_{1},
\label{condpred} \\
\text{ \ }\bar{c}_{1}& :=\frac{2\alpha c_{1}+c_{0}c_{2}+c_{2}^{2}}{c_{2}^{2}}%
,\text{ \ }\bar{c}_{2}:=\frac{2\alpha }{c_{2}^{2}}.  \label{nno}
\end{align}
All the constants in (\ref{condpred2}) are positive by construction and $\gamma$ and $\tau$ are nonnegative by definition. Then, a necessary condition for (\ref{condpred2}) to be satisfied is $\lambda_2 \gamma> \gamma^{\prime} $. After some straightforward computations, inequality (\ref{condpred2}) can be rewritten as follows
\begin{equation}
\tau <-\Big(\bar{c}_{2}+\frac{\bar{c}_{1}}{\gamma \lambda _{k}}\Big)\pm
\sqrt{\Big(\bar{c}_{2}+\frac{\bar{c}_{1}}{\gamma \lambda _{k}}\Big)^{2}+%
\frac{2\bar{c}_{2}(\lambda _{2}\gamma -\gamma ^{\prime })}{\lambda
_{k}^{2}\gamma ^{2}}}.  \label{neww1}
\end{equation}
The time-delay $\tau$ is nonnegative by definition. Hence, in order to satisfy (\ref{neww1}), it is sufficient to consider the possible positive values of the right-hand side of (\ref{neww1}), i.e., the positive square root. Then, inequality (\ref{neww1}) boils down to
\begin{align} \label{neww2}
\left\{ \begin{array}{l}
\tau <\varphi (\gamma ),\\
\varphi (\gamma ):=-\Big(\bar{c}_{2}+\frac{\bar{c}_{1}}{\gamma \lambda
_{k}}\Big)+\sqrt{\Big(\bar{c}_{2}+\frac{\bar{c}_{1}}{\gamma \lambda _{k}}%
\Big)^{2}+\frac{2\bar{c}_{2} (\lambda _{2} \gamma -\gamma ^{\prime })}{\lambda
_{k}^{2}\gamma ^{2}}}.
\end{array} \right.
\end{align}
We are only interested in possible values of $\gamma,\tau \in \Real_{\geq 0}$ such that (\ref{neww2}) is satisfied. Then, we restrict the function $\varphi(\gamma)$ to the set $\mathcal{J}:=[\frac{\gamma^\prime}{\lambda_2},\infty)$. Next, we prove that the function $\varphi:\mathcal{J} \rightarrow \Real_{\geq 0}$ is unimodal. The function $\varphi(\cdot)$ is continuous and real-valued on $\mathcal{J}$. Moreover, it is strictly positive on the interior of $\mathcal{J}$, it has a root at $\gamma = \frac{\gamma^\prime}{\lambda_2}$, i.e., $\varphi(\frac{\gamma^\prime}{\lambda_2})=0$, and $\lim_{\gamma \rightarrow \infty }\varphi (\gamma )$ equals
\begin{align*}
&\lim_{\gamma
\rightarrow \infty }\frac{\allowbreak \frac{2\bar{c}_{2}}{%
\lambda _{k}^{2}}\left( \frac{\lambda _{2}}{\gamma }-\frac{\gamma ^{\prime }}{\gamma
^{2}}\right) }{\left( \bar{c}_{2}+\frac{\bar{c}_{1}}{\gamma \lambda _{k}}%
\right) +\sqrt{(\bar{c}_{2}+\frac{\bar{c}_{1}}{\gamma \lambda _{k}})^{2}+%
\frac{2\bar{c}_{2}}{\lambda _{k}^{2}}\left( \frac{\lambda _{2}}{\gamma }-%
\frac{\gamma ^{\prime }}{\gamma ^{2}}\right) }} \\[1mm]
&=\frac{\frac{2\bar{c}_{2} }{\lambda _{k}^{2}}\left( 0\right) }{%
\left( \bar{c}_{2}+0\right) +\sqrt{(\bar{c}_{2}+0)^{2}+\frac{2\bar{c}%
_{2}}{\lambda _{k}^{2}}\left( 0\right) }}=\frac{0}{\allowbreak 2%
\bar{c}_{2}}=0.
\end{align*}
The function $\varphi(\cdot)$ is differentiable on $\mathcal{J}$, then we can compute its local extrema by computing its critical points. It is easy to verify that $\frac{\partial \varphi (\gamma )}{\partial \gamma }=0$ only for $\gamma = \gamma^*$ and $\gamma = \tilde{\gamma}$ with
\begingroup\makeatletter\def\f@size{7.5}\check@mathfonts
\def\maketag@@@#1{\hbox{\m@th\normalsize\normalfont#1}}%
\begin{align}
\gamma ^{\ast }& =\left( 1+\frac{\lambda _{2}}{\lambda _{2}+2\lambda _{k}%
\bar{c}_{1}}\right) \frac{\gamma ^{\prime }}{\lambda _{2}}+\frac{\sqrt{2\bar{%
c}_{1}^{2}\bar{c}_{2}\gamma ^{\prime }(\lambda _{2}^{2}+2\lambda _{2}\lambda
_{k}\bar{c}_{1}+2\lambda _{k}^{2}\bar{c}_{2}\gamma ^{\prime })}}{\bar{c}%
_{2}\lambda _{2}(\lambda _{2}+2\lambda _{k}\bar{c}_{1})},  \label{root1}\\
\tilde{\gamma}\text{ }& =\left( 1+\frac{\lambda _{2}}{\lambda _{2}+2\lambda
_{k}\bar{c}_{1}}\right) \frac{\gamma ^{\prime }}{\lambda _{2}}-\frac{\sqrt{2%
\bar{c}_{1}^{2}\bar{c}_{2}\gamma ^{\prime }(\lambda _{2}^{2}+2\lambda
_{2}\lambda _{k}\bar{c}_{1}+2\lambda _{k}^{2}\bar{c}_{2}\gamma ^{\prime })}}{%
\bar{c}_{2}\lambda _{2}(\lambda _{2}+2\lambda _{k}\bar{c}_{1})}.
\label{root2}
\end{align}\endgroup
Then, $\gamma = \gamma^*$ and $\gamma = \tilde{\gamma}$ are the critical points of $\varphi(\gamma)$ and $\varphi(\gamma^*)$ and $\varphi(\tilde{\gamma})$ are the corresponding global extrema. Notice that $\gamma^{\ast} > \frac{\gamma^\prime}{\lambda_2}$; therefore, $\gamma^{\ast}$ belongs to the interior of $\mathcal{J}$. It is difficult to visualize from (\ref{root2}) whether $\tilde{\gamma}$ is contained in $\mathcal{J}$. Then, we rewrite (\ref{root2}) in a more suitable manner
\begin{equation}
\tilde{\gamma}=\frac{2\bar{c}_{2}\gamma ^{\prime }-\bar{c}_{1}^{2}}{\bar{c}%
_{2}\left( \lambda _{2}+\bar{c}_{1}\lambda _{k}\right) +\bar{c}_{1}\sqrt{%
\frac{\bar{c}_{2}(\lambda _{2}^{2}+2\lambda _{2}\lambda _{k}\bar{c}%
_{1}+2\lambda _{k}^{2}\bar{c}_{2}\gamma ^{\prime })}{2\gamma ^{\prime }}}}.
\label{root2p}
\end{equation}
Note that the denominator of (\ref{root2p}) is strictly positive, then the sign of $\tilde{\gamma}$ is solely determined by the numerator. Substitution of (\ref{condpred}) and (\ref{nno}) in the numerator of (\ref{root2p}) yields
\begin{equation}
2\bar{c}_{2}\gamma^{\prime }-\bar{c}_{1}^{2}=-\frac{4\alpha c_{1}\left(
c_{0}c_{2}+\alpha c_{1}\right) }{c_{2}^{2}},  \label{root3p}
\end{equation}
which is strictly negative. It follows that  $\tilde{\gamma}$ is strictly negative as well; in consequence, $\tilde{\gamma}$ is not contained in $\mathcal{J}$, i.e., $\tilde{\gamma} \notin \mathcal{J}$. Then, the function $\varphi(\cdot)$ has a unique extremum on $\mathcal{J}$ and it is given by $\varphi(\gamma^*)$. Moreover, given that $\varphi(\gamma^\prime)=0$, $\lim_{\gamma \rightarrow \infty }\varphi (\gamma)=0$, $\phi(\gamma)$ is strictly positive on the interior of $\mathcal{J}$, and $\varphi(\gamma^*)$ is the unique extremum on $\mathcal{J}$, it follows that $\varphi(\gamma^*)$ is a unique local maximum on $\mathcal{J}$; therefore, it can be concluded that the function $\varphi(\cdot)$ is a \emph{unimodal function} in the sense of Definition 3. Hence, (\ref{102AT3}) is negative definite if $\lambda_2 \gamma > \gamma^\prime$ and $\tau < \varphi(\gamma)$. Finally, ultimate boundedness of the solutions and the Lyapunov-Razumikhin theorem imply that the set $\{\bar{\zeta}=\bar{y}=0 \}$ is a global attractor for $\lambda_2 \gamma > \gamma^\prime$ and $\tau < \varphi(\gamma)$.\hfill $\blacksquare$

\subsection{Proof of Corollary 2}

From Theorem 1, the optimal coupling strength $\gamma^*$ for any strongly connected graph is given by (\ref{root1}). Moreover, the corresponding maximum time-delay $\tau^*$ is given by $\varphi(\gamma^*)$ with unimodal function $\varphi(\cdot)$ given in (\ref{neww2}). Then, after some straightforward algebra, $\gamma^*$ and $\tau^*= \varphi (\gamma ^{\ast })$ can be written as follows
\begin{align}
\gamma ^{\ast } =\frac{\gamma ^{\prime }}{\lambda _{2}} &  \vast( 1+\frac{1}{1+2%
\bar{c}_{1}\frac{\lambda _{k}}{\lambda _{2}}}   \notag \\%
[0.16in]
 &\text{ \ }  +\sqrt{1+\frac{1}{\left( 1+2\bar{c}_{1}\frac{\lambda _{k}}{\lambda
_{2}}\right) ^{2}}+\frac{2\left( \bar{c}_{1}^{2}-\bar{c}_{2}\gamma ^{\prime
}\right) }{\bar{c}_{2}\gamma ^{\prime }\left( 1+2\bar{c}_{1}\frac{\lambda
_{k}}{\lambda _{2}}\right) }}\vast) , \label{gammaS}
\end{align}
\begin{equation}
\tau ^{\ast } =\frac{\bar{c}_{2}}{\frac{\lambda _{k}}{\lambda _{2}} \left(
\bar{c}_{1}+2\bar{c}_{2}\gamma ^{\prime }\frac{\lambda _{k}}{\lambda _{2}}+%
\sqrt{2\bar{c}_{2}\gamma ^{\prime }+4\bar{c}_{2}\bar{c}_{1}\gamma ^{\prime }%
\frac{\lambda _{k}}{\lambda _{2}}+\left( 2\bar{c}_{2}\gamma ^{\prime }\frac{%
\lambda _{k}}{\lambda _{2}}\right) ^{2}}\right) }.  \label{TauS}
\end{equation}
Next, we analyze the cases stated in Corollary 2. Case (a): The maximum time-delay in (\ref{TauS}) relies on the constants $\bar{c}_1,\bar{c}_2,\gamma^\prime \in \Real_{>0}$ and the quotient $\frac{\lambda_k}{\lambda_2}$. The constants $\bar{c}_1,\bar{c}_2$, and $\gamma^\prime$ do not depend on the network topology (they depend on the vectorfields $q(\cdot)$ and $a(\cdot)$), then the effect of network is solely determined by $\frac{\lambda_k}{\lambda_2}$. From Gerschgorin's disc theorem, it can be concluded that the eigenvalues $\lambda_2$ and $\lambda_k$ of any strongly connected undirected graph are positive real, i.e., $0 < \lambda_2 \leq \lambda_k$, and therefore $\frac{\lambda_k}{\lambda_2} \geq 1$. Then, it is clear from (\ref{TauS}) that networks with equal $\frac{\lambda_k}{\lambda_2}$ have the same maximum time-delay. Moreover, from (\ref{gammaS}), it is clear that if two networks have the same quotient $\frac{\lambda_k}{\lambda_2}$, then the value of $\gamma^*$ is solely determined by $\lambda_2$, the larger the $\lambda_2$ the smaller the $\gamma^*$, and vice versa. Case (b): The eigenvalues $\lambda_2$ and $\lambda_k$ of any undirected strongly connected graph are positive real and $0 < \lambda_2 \leq \lambda_k$; therefore, $\frac{\lambda_k}{\lambda_2} \geq 1$. From (\ref{TauS}), it is clear that $\tau^*$ has its maximum value at $\frac{\lambda_k}{\lambda_2}=1$. The part regarding $\gamma^*$ follows from the same arguments of case (a). Case (c): Clearly, from (\ref{TauS}), the larger the quotient $\frac{\lambda_k}{\lambda_2}$ the smaller the $\tau^*$, and vice versa. Moreover, from (\ref{gammaS}), it is clear that the larger the quotient $\frac{\lambda_k}{\lambda_2}$ the smaller the $\gamma^*$ for a fixed $\lambda_2$, and for a fixed quotient, the larger the $\lambda_2$ the smaller the $\gamma^*$ and the assertion follows. \hfill $\blacksquare$

\bibliographystyle{plain}
\bibliography{ifacconf2}

%
%
%
%
%
%
%
%
%
%
%
%
%
%
%
%
%
%
%

\end{document}